\documentclass[%
 aip,jcp,
 amsmath,amssymb,
 reprint,
]{revtex4-1}

\usepackage{graphicx}
\usepackage{dcolumn}
\usepackage{bm}

\usepackage[utf8]{inputenc}
\usepackage[T1]{fontenc}
\usepackage{mathptmx}
\usepackage{listings}
\usepackage{rotating} 

\usepackage{color}
\usepackage{dcolumn} 
\usepackage{bm} 
\usepackage{graphicx}
\usepackage{multirow} 
\usepackage{pifont} 
\usepackage{epsfig}
\usepackage{amsmath} 
\usepackage{subfigure}
\usepackage{float}
\usepackage{booktabs}
\usepackage{tabularx}
\usepackage{natbib}
\usepackage{gensymb}
\usepackage{rotating}
\usepackage[normalem]{ulem}

\begin{document}

\author{Marcus D. Liebenthal}
\affiliation{
             Department of Chemistry and Biochemistry,
             Florida State University,
             Tallahassee, FL 32306-4390}
\author{Nam Vu}
\affiliation{
             Department of Chemistry and Biochemistry,
             Florida State University,
             Tallahassee, FL 32306-4390}
             
\author{A. Eugene DePrince III}
\email{adeprince@fsu.edu}
\affiliation{
             Department of Chemistry and Biochemistry,
             Florida State University,
             Tallahassee, FL 32306-4390}

\title{Equation-of-motion cavity quantum electrodynamics coupled-cluster theory for electron attachment}

\begin{abstract}

The electron attachment variant of equation-of-motion coupled-cluster theory (EOM-EA-CC) is generalized to the case of strong light-matter coupling within the framework of cavity quantum electrodynamics (QED). The resulting EOM-EA-QED-CC formalism provides an {\em ab initio}, correlated, and non-perturbative description of cavity-induced effects in many-electron systems that complements other recently proposed cavity-QED-based extensions of coupled-cluster theory. Importantly, this work demonstrates that QED generalizations of EOM-CC theory are useful frameworks for exploring particle non-conserving sectors of Fock space, thereby establishing a path forward for the simultaneous description of both strong electron-electron and electron-photon correlation effects.

\end{abstract}

\maketitle

\section{Introduction}

Strong coupling of photonic and molecular degrees of freedom can lead to the formation of hybrid light-matter states known as polaritons that can exhibit dramatically different properties relative to the original uncoupled states.\cite{Nori19_19,Ebbesen16_2403} Recent improvements to optical cavities have enabled such strong coupling even in the single-molecule limit and at room temperature.\cite{Baumberg16_127} From a chemist’s or materials scientist’s perspective, a fundamental understanding of the principles that govern polariton formation and manipulation is desirable, as such knowledge can facilitate the realization of light-mediated control over chemical reactivity,\cite{Shalabney21, Bucksbaum15_164003, Johannes19_131, Zhou19_4685} stereo/regioselectivity,\cite{Ebbesen2019_615,Corni18_4688} or even superconductive properties.\cite{Rubio18_6969}
Accordingly, experimental advances in strong light-matter coupling must be accompanied by  predictive theoretical frameworks that can explain the chemical consequences of polariton formation and assist in the development of design principles for polariton-based chemical transformations.

Most theoretical descriptions of strong light-matter coupling involve model Hamiltonians that capture interactions between few-level quantum emitters and the quantized electromagnetic field of an optical cavity.\cite{JaynesCummings63_89, TavisCummings1968_379, Chiao08_716, Aspect2010, Becker06_1325} These models can provide critical insights into the properties of polaritons and have been used in a large number of theoretical studies describing cavity-mediated changes to energy landscapes and the resulting effects on energy transfer and reactivity. \cite{Corni18_4688,Zhou19_4685,Zhou18_6325,Huo19_5519,Huo20_6321,Foley20_9063,Genes19_203602} 
Even greater insights and predictivity can be achieved via {\it ab initio} cavity quantum electrodynamics (QED) approaches\cite{Bauer11_042107,Rubio14_012508,Tokatly13_233001,Rubio15_093001,Rubio17_3026,Rubio18_992,Appel19_225,Narang20_094116,Rubio18_arxiv,Rubio19_2694,Rubio20_5601,Shao21_064107,Koch20_041043, Manby20_023262, Corni21_6664,Koch21_094113,DePrince21_094112, Flick21_9100,Chan20_224112,Flick21_arxiv_proton_transfer} that describe electron--photon interactions from first principles. The majority of such studies have employed a cavity QED generalization of density functional theory (DFT),\cite{Bauer11_042107,Rubio14_012508,Tokatly13_233001,Rubio17_3026,Tokatly18_235123,Rubio15_093001,Rubio18_992,Appel19_225,Narang20_094116,Shao21_064107} which provides an economical description of electron correlation effects and can capture qualitative changes induced via strong coupling to light. Nevertheless, the application of cavity QED extensions of DFT to general systems is potentially hampered by the well-known problems that plague standard formulations of DFT\cite{Yang08_792} and the relatively small number of available exchange--correlation functionals for the polaritonic problem.\cite{Rubio15_093001,Rubio18_992,Flick21_arxiv} 

A reliable and systematically-improvable {\it ab initio} description of electron-electron and electron-photon correlation can be realized via many-body approaches such as coupled-cluster (CC) theory\cite{Cizek66_4256, Paldus71_359, Bartlett09_book, Musial07_291, Bartlett07_291} and excited-state generalizations of CC theory based on the equation of motion (EOM)\cite{Bartlett93_7029,Bartlett12_126,Musial07_291,Krylov08_433} or closely-related linear response\cite{Monkhorst77_421,Mukherjee79_325,Monkhorst83_1217,Jorgensen90_3333,Helgaker90_3345,Koch97_8059,Jorgensen95_7429,Jorgensen19_134109} formalisms. Koch and coworkers\cite{Koch20_041043} recently developed a cavity-QED-based extension of CC theory that has since been applied to the description of cavity-induced changes to both ground-\cite{Koch20_041043,DePrince21_094112,Koch21_094113,Flick21_9100,Flick21_arxiv_proton_transfer} and excited-state\cite{Koch20_041043,Corni21_6664,Flick21_9100} properties of coupled electron-photon systems using fully {\em ab initio} Hamiltonians. A similar extension of CC theory using slightly different excitation operators within the photon space and a model Hamiltonian has also been realized.\cite{Manby20_023262} These studies have established QED-CC theory as a useful tool for the description of strongly-coupled light-matter systems and have paved the way for additional theoretical developments that broaden QED-CC's application domain.

To date, all prior QED-CC-based descriptions of excited-state energies and properties have been obtained using a particle-conserving {\em ansatz} similar to the excitation-energy (EE) formulation of EOM-CC theory (EOM-EE-CC).\cite{Bartlett89_57,Bartlett95_81,Bartlett94_3073,Piecuch00_8490,Piecuch01_643} However, there is no fundamental reason why QED-CC  cannot be generalized such that the excitation operators can access different ({\em i.e.}, particle non-conserving) sectors of Fock space, as is done, for example, in the electron attachment (EA)\cite{Bartlett95_3629,Bartlett95_6735,Wloch05_134113,Piecuch06_234107,Hirata07_134112} or ionization potential (IP)\cite{Snijders92_55,Snijders93_15,Gauss94_8938,Bartlett03_1128,Wloch05_134113,Piecuch06_234107,Krylov11_6028,Krylov12_2726} formulations of EOM-CC  (EOM-EA-CC and EOM-IP-CC, respectively). In this work, we consider the case of electron attachment for molecules strongly-coupled {\color{black}to} the quantized electromagnetic field of an optical cavity and develop the QED-CC analogue of EOM-EA-CC theory. This EOM-EA-QED-CC approach can be used to determine electron affinities of cavity-bound molecules or, more generally, to compute the polaritonic spectrum of an ($N$+1)-electron polaritonic state, starting from an $N$-electron one. The latter use case could prove useful when seeking excited-state information for systems with difficult-to-describe ground states. We consider both applications of the approach and demonstrate that (i) electron affinities from EOM-EA-QED-CC are in good agreement with those obtained from QED-CC calculations on different charge states and (ii) EOM-EA-QED-CC captures expected qualitative features associated with strong light-matter coupling in a cavity-bound molecule.

\section{Theory}

Consider a molecule or collection of molecules embedded in a cavity that supports a single optical mode.  Interactions between the molecular degrees of freedom and the quantized electromagnetic field of the cavity can be described by the Pauli--Fierz Hamiltonian,\cite{Spohn04_book,Rubio18_0118} which, in the length gauge and within the dipole and Born-Oppenheimer approximations, takes the following form:
\begin{eqnarray}
    \label{EQN:PFH_COHERENT}
    \hat{H} &=& \hat{H}_{\rm e} + \omega_{\rm cav} \hat{b}^\dagger \hat{b} - ( \omega_{\rm cav}/2 )^{1/2} ({\bm \lambda} \cdot [{\bm \mu} - \langle {\bm \mu }\rangle] )(\hat{b}^\dagger +\hat{b}) \nonumber \\
    &+& \frac{1}{2} ({\bm \lambda} \cdot [{\bm \mu} - \langle {\bm \mu }\rangle] )^2
\end{eqnarray}
Here, $\hat{H}_{\rm e}$ represents the electronic Hamiltonian that arises in standard electronic structure theories. The second term represents the Hamiltonian for the cavity mode, which is a harmonic oscillator with fundamental frequency $\omega_{\rm cav}$ and coupling vector ${\bm \lambda}$, and the symbols $\hat{b}^\dagger$ and $\hat{b}$ are bosonic creation and annihilation operators, respectively. The last two terms are the bilinear coupling and dipole self-energy terms, respectively, and the symbols ${\bm \mu}$ and $\langle {\bm \mu }\rangle$ represent the molecular dipole operator and the expectation value of this operator with respect to a mean-field ({\em i.e.}, QED Hartree-Fock [HF]) wave function. Note that the usual Pauli--Fierz Hamiltonian lacks the dipole expectation values; these terms have been introduced as a matter of convenience through the coherent-state transformation described in Ref.~\citenum{Koch20_041043}.

We model the ground state of the cavity-embedded molecule using the QED-CC formalism outlined in Ref.~\citenum{Koch20_041043} in which the wave function is defined as
\begin{equation}
    |\Psi_{\rm CC}\rangle = e^{\hat{T}}|0^{\rm e}0^{\rm p}\rangle
\end{equation}
Here, the superscripts ``e'' and ``p'' refer to electronic and photonic degrees of freedom, respectively, and the QED-HF reference wave function, $|0^{\rm e}0^{\rm p}\rangle$, is defined as a direct product of a determinant of electronic orbitals, $|0^{\rm e}\rangle$, and a zero-photon photon-number state, $|0^{\rm p}\rangle$. The electronic component of the reference function can be determined via a standard Hartree-Fock-Roothan procedure using the Hamiltonian in Eq.~\ref{EQN:PFH_COHERENT}; after integrating out the photon degrees of freedom, only $\hat{H}_{\rm e}$ and the dipole self-energy term enter this procedure. For QED-CC with single and double electron excitations and up to single photon transitions (QED-CCSD-1),\cite{Koch20_041043} the cluster operator, $\hat{T}$ is defined as
\begin{eqnarray}
\label{EQN:T}
    \hat{T} &=& \sum_{ia} t_i^a \hat{a}^\dagger_a \hat{a}_i  + \frac{1}{4} \sum_{ijab} t_{ij}^{ab} \hat{a}^\dagger_a \hat{a}^\dagger_b \hat{a}_j \hat{a}_i \nonumber \\
    &+& \sum_{ia} u_i^a \hat{a}^\dagger_a \hat{a}_i \hat{b}^\dagger + \frac{1}{4} \sum_{ijab} u_{ij}^{ab} \hat{a}^\dagger_a \hat{a}^\dagger_b \hat{a}_j \hat{a}_i \hat{b}^\dagger \nonumber \\
    &+& u \hat{b}^\dagger
\end{eqnarray}
The symbols $\hat{a}^\dagger$ and $\hat{a}$ represent fermionic creation and annihilation operators, respectively, and the labels $i$~/~$j$ and $a$~/~$b$ refer to orbitals that are occupied or unoccupied in the QED-HF reference function. Given Eq.~\ref{EQN:T}, we see that QED-CCSD-1 is simply a generalization of standard CC with single and double excitations (CCSD)\cite{Bartlett82_1910} for the coupled electron-photon case that includes up to single photon transitions in the cluster operator. The cluster amplitudes, $t_i^a$, $t_{ij}^{ab}$, $u_i^a$, $u_{ij}^{ab}$, and $u$ can be determined in the usual projective way, {\em i.e.} by solving the equations, 
\begin{eqnarray}
\label{EQN:CC_ENERGY}
\langle 0^{\rm e} 0^{\rm p}  | e^{-\hat{T}} \hat{H} e^{\hat{T}}|0^{\rm e}0^{\rm p}\rangle  &=& E_{\rm
CC}  \\
\label{EQN:CC_AMPS}
\langle \mu^{\rm e} \nu^{\rm p}  | e^{-\hat{T}} \hat{H} e^{\hat{T}}|0^{\rm e}0^{\rm p}\rangle  &=& 0  
\end{eqnarray}
where $E_{\rm CC}$ is the energy associated with the ground-state wave function. Here, $\langle \mu^{\rm e} \nu^{\rm p}|$ refers to a state defined by the direct product of a determinant of electronic orbitals, $\langle \mu^{\rm e} |$, and a photon number state, $\langle \nu^{\rm p}|$. At the QED-CCSD-1 level of theory, $\langle \mu^{\rm e} |$ could refer to the reference electronic configuration or any singly / doubly substituted configuration, and $\langle \nu^{\rm p}|$ refers to one of only two photon states, representing either zero or one photon in the cavity. Given these possibilities, we note that the $\langle \mu^{\rm e} \nu^{\rm p}| = \langle 0^{\rm e} 0^{\rm p}  |$ case is excluded from Eq.~\ref{EQN:CC_AMPS}.

Once one has determined the amplitudes that define the ground-state $N$-electron wave function, $|\Psi_{\rm CC}\rangle$, excited-states can be parametrized using the EOM-CC formalism\cite{Bartlett93_7029}
\begin{eqnarray}
\langle \Psi_I | = \langle 0^{\rm e}0^{\rm p} | e^{-\hat{T}} \hat{L}_I \\
| \Psi_I \rangle = \hat{R}_I e^{\hat{T}} |  0^{\rm e}0^{\rm p} \rangle 
\end{eqnarray}
where the label $I$ denotes different states. We are interested in the case of electron attachment, so $\hat{R}_I$ and $\hat{L}_I$ should represent particle-non-conserving transition operators that, at the EOM-EA-QED-CCSD-1 level of theory, take the form
\begin{eqnarray}
        \hat{R}_I &=&\sum_{a} r^a \hat{a}^\dagger_a + \frac{1}{2} \sum_{abi} r_{i}^{ab} \hat{a}^\dagger_a \hat{a}^\dagger_b \hat{a}_i \nonumber \\
        &+& \sum_{a} s^a \hat{a}^\dagger_a \hat{b}^\dagger + \frac{1}{2} \sum_{abi} s_{i}^{ab} \hat{a}^\dagger_a \hat{a}^\dagger_b \hat{a}_i \hat{b}^\dagger,
\end{eqnarray}
and
\begin{eqnarray}
        \hat{L}_I &=&\sum_{a} l_a \hat{a}_a + \frac{1}{2} \sum_{abi} l^{i}_{ab}  \hat{a}^\dagger_i \hat{a}_b \hat{a}_a \nonumber \\
        &+& \sum_{a} m_a \hat{a}_a \hat{b} + \frac{1}{2} \sum_{abi} m^{i}_{ab} \hat{a}^\dagger_i \hat{a}_b \hat{a}_a  \hat{b},
\end{eqnarray}
respectively. Like QED-CCSD-1, EOM-EA-QED-CCSD-1 is a generalization of standard EOM-EA-CCSD for the coupled electron-photon case that includes up to single photon transitions in the right- and left-hand transition operators. The transition amplitudes are then determined by solving right- and left-hand eigenvalue equations
\begin{equation}
    \bar{H} \hat{R}_I |  0^{\rm e}0^{\rm p} \rangle = E_I \hat{R}_I |  0^{\rm e}0^{\rm p} \rangle 
\end{equation}
and
\begin{equation}
    \langle  0^{\rm e}0^{\rm p} |  \hat{L}_I \bar{H}  =  \langle 0^{\rm e}0^{\rm p}  | \hat{L}_I E_I
\end{equation}
respectively, where $\bar{H} = e^{-\hat{T}}\hat{H}e^{\hat{T}}$ is the similarity-transformed Pauli--Fierz Hamiltonian, and $E_I$ represents the energy of the $I^{th}$ ($N$+1)-electron state. If EAs are desired, they can be evaluated as the difference between $E_I$ and $E_{\rm CC}$. 

\section{Computational Details}

The EOM-EA-QED-CCSD-1 model was implemented as a plugin to the \textsc{Psi4} electronic structure package.\cite{Sherrill20_184108} The working equations for the ground-state QED-CCSD-1 and the EOM-EA-QED-CCSD-1 approaches were generated using a locally-modified version of \texttt{p$^\dagger$q},\cite{DePrince21_e1954709} which is a library for manipulating strings of second-quantized operators such as those that arise in coupled-cluster theory.  We also report some results evaluated using the excitation energy (EE)\cite{Bartlett89_57,Bartlett95_81,Bartlett94_3073,Piecuch00_8490,Piecuch01_643} form of EOM-CCSD (EOM-EE-CCSD); all EOM-EE-CCSD calculations were carried out using the implementation in \textsc{Psi4}. 

Following Ref.~\citenum{DePrince21_094112}, calculations involving sodium halide compounds (NaX, X = F, Cl, Br, I) were carried out in the def2-TZVPPD basis set, using the density-fitting approximation to the electron repulsion integral (ERI) tensor.\cite{Whitten73_4496,Sabin79_3396}  The def2-universal-JKFIT and def2-TZVPPD-RIFIT auxiliary basis sets were used in the QED-HF and (EOM-)QED-CC portions of the calculations, respectively. Geometries for isolated neutral NaX were optimized using {\color{black}DFT} theory, with the B3LYP exchange-correlation functional, using the same primary and auxiliary basis sets used in the corresponding QED-HF calculations. It should also be noted that calculations on NaI used the def2-TZVPPD effective core potential for iodine. Calculations involving magnesium fluoride (MgF) and MgF$^{+}$ were carried out in the aug-cc-pVDZ basis set. EOM-EA-QED-CC calculations applied to MgF$^+$ employed Cholesky-decomposed ERIs with a tight threshold of 1$\times 10^{-12}$ E$_{\rm h}$, while EOM-EE-CC calculations on neutral MgF used exact four-center ERIs.

 {\color{black}Before exploring the numerical properties of EOM-EA-QED-CCSD-1 in the next section, we comment briefly on the computational cost associated with this method and related QED extensions of EOM-CC theory. The construction of a single $\sigma$-vector (the action of the similarity-transformed Hamiltonian or its conjugate transpose on a trial solution vector) within the standard EOM-EE- or EOM-EA-CCSD approaches is dominated by steps with floating-point costs of $\mathcal{O}(k^6)$ and $\mathcal{O}(k^5)$, respectively, where $k$ is a measure of the size of the system. The formal scalings of these operations within QED generalizations of EOM-EE- and EOM-EA-CC are unchanged, but, when considering single photon transitions, the prefactors roughly double. For example, assuming that the number of occupied spin-orbitals, $o$, is much less than the number of virtual spin-orbitals, $v$, the construction of the right-hand $\sigma$-vector in EOM-EA-CC is dominated by a single $\mathcal{O}(ov^4)$ tensor contraction. At the EOM-EA-QED-CCSD-1 level of theory, there are two such contractions. Should one choose to include additional photon transitions, the cost of the construction of the $\sigma$-vector should increase roughly linearly with the photon excitation rank.}

\section{Results and Discussion}

\begin{figure}[!htpb]
    \centering
    \includegraphics{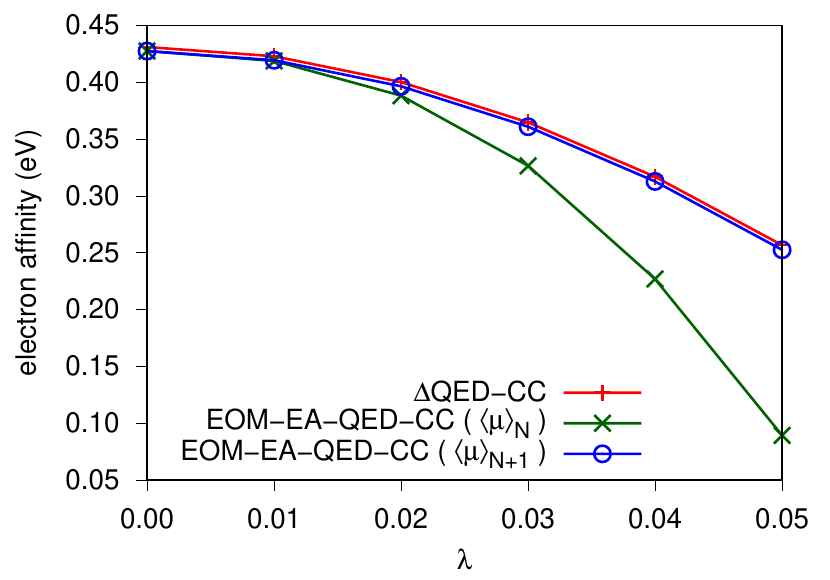}
    \caption{EOM-EA-QED-CCSD-1 and $\Delta$QED-CCSD-1 electron affinities (eV) for sodium fluoride as a function of the cavity coupling strength, $\lambda$.}
    \label{FIG:DELTA_VERSUS_EOM}
\end{figure}

We begin by evaluating electron affinities for the series of sodium halide compounds (NaX, X = F, Cl, Br, I) considered in Ref.~\citenum{DePrince21_094112}. In that work, $\Delta$QED-CCSD-1 electron affinities were computed via energy differences between the charge states, {\em i.e.}, 
\begin{equation}
{\rm EA} = E(N) - E(N+1)
\end{equation}
where $E(N)$ and $E(N+1)$ represent the energy of the neutral and anion species, respectively. Figure \ref{FIG:DELTA_VERSUS_EOM} depicts EOM-EA-QED-CCSD-1 and $\Delta$QED-CCSD-1 electron affinities for one of these compounds (NaF) when coupled to a single-mode optical cavity with a resonant frequency $\omega_{\rm cav}$ = 2.0 eV and the cavity mode polarized along the $z$-axis [${\bm \lambda} = (0, 0, \lambda)$]. {\color{black}
We can relate this coupling strength to the effective mode volume for the cavity as\cite{Feist19_021057}
\begin{equation}
    \lambda = \frac{1}{\sqrt{\epsilon_0 V_{\rm eff}}}
\end{equation}
The largest coupling strength that we consider, $\lambda$ = 0.05, corresponds to $V_{\rm eff}$ $\approx$ 0.74 nm$^3$, which is in line with the sub-nm$^3$ effective volumes demonstrated experimentally in what have been termed optical ``picocavities.''\cite{Baumberg16_726,Baumberg18_7146} }
The molecular axis is chosen to be parallel to the cavity mode axis. 

As reported in Ref.~\citenum{DePrince21_094112}, electron affinities for this compound decrease with increasing coupling strengths. We find that the naive application of EOM-EA-QED-CCSD-1 can reproduce this qualitative trend, but quantitative agreement with $\Delta$QED-CCSD-1 is poor. Consider the green curve in Fig.~\ref{FIG:DELTA_VERSUS_EOM}; these data were generated via EOM-EA-QED-CCSD-1 using the Hamiltonian presented in Eq.~\ref{EQN:PFH_COHERENT}, which depends on the average dipole moment associated with the QED-HF wave function for neutral NaF (hence the label in Fig.~\ref{FIG:DELTA_VERSUS_EOM}, $\langle\mu\rangle_{\rm N}$). The resulting electron affinities differ significantly from those from $\Delta$QED-CCSD-1, particularly at large coupling strengths. For example, at $\lambda = 0.05$, electron affinities derived from $\Delta$QED-CCSD-1 (0.26 eV) and EOM-EA-QED-CCSD-1 (0.09 eV) differ by 0.17 eV.

The somewhat surprisingly poor performance of EOM-EA-QED-CCSD-1 stems from the coherent-state-basis transformation of the Hamiltonian, which introduces a reference-state dependence into this operator. Specifically, the bilinear coupling and dipole self-energy terms depend explicitly on the expectation value of the dipole operator with respect to the QED-HF wave function. Consequently, $\Delta$QED-CCSD-1 derived electron affinities can be expected to deviate from those from EOM-EA-QED-CCSD-1 if the QED-HF dipole moments associated with the $N$- and ($N$+1)-electron states differ. It therefore seems reasonable that the coherent-state transformation applied to the Hamiltonian used in the EOM-EA-QED-CCSD-1 procedure should be defined by the dipole moment associated with the QED-HF wave function for the ($N$+1)-electron state. Indeed, any EOM-based QED-CC calculations that explore particle/spin-non-conserving sectors of Fock space should use the dipole moment associated with the QED-HF wave function for the target manifold of states to define the coherent-state basis used in the EOM-CC procedure. Electron affinities computed via EOM-EA-QED-CCSD-1 with the coherent-state-basis transformation defined by the anion reference are depicted in Fig.~\ref{FIG:DELTA_VERSUS_EOM} (the blue curve labeled by $\langle\mu\rangle_{\rm N+1}$). With the correct choice for the coherent-state basis, we find that EOM-EA-QED-CCSD-1 closely reproduces electron affinities from the $\Delta$QED-CCSD-1 approach. The reasonableness of this choice for the coherent-state basis is further validated by the electron affinities for the other members of the sodium halide series, which are tabulated in Table \ref{TAB:NAX}. We find that, for these compounds, EOM-EA-QED-CCSD-1 electron affinities deviate from those obtained via $\Delta$QED-CCSD-1 by, at worst, 0.01 eV over the range of coupling strengths considered. 

\begin{table}[!htpb]
    \caption{EOM-EA-QED-CCSD-1 and $\Delta$QED-CCSD-1 electron affinities (eV) for sodium halide compounds embedded within an optical cavity with a fundamental frequency $\omega$ = 2.0 eV. The cavity mode polarization is parallel to the molecular axis. $\Delta$QED-CCSD-1 results are taken from Ref.~\citenum{DePrince21_094112}.}
    \label{TAB:NAX}
    \begin{center}
        \begin{tabular}{cccccccccc}

            \hline\hline
        
  	        & \multicolumn{4}{c}{$\Delta$QED-CC} &~~& \multicolumn{4}{c}{EOM-EA-QED-CC} \\	
			\cline{2-5} \cline{7-10}					
      $\lambda$    	&NaF	&NaCl	&NaBr	&NaI	&&NaF	&NaCl	&NaBr	&NaI    \\
      	\hline
0.00	&0.43	&0.64	&0.70	&0.78	&& 0.43 & 0.65 & 0.71 & 0.78 \\
0.01	&0.42	&0.64	&0.70	&0.77	&& 0.42 & 0.64 & 0.70 & 0.78 \\
0.02	&0.40	&0.62	&0.68	&0.75	&& 0.40 & 0.62 & 0.68 & 0.76 \\
0.03	&0.36	&0.59	&0.65	&0.72	&& 0.36 & 0.59 & 0.65 & 0.73 \\
0.04	&0.32	&0.54	&0.60	&0.68	&& 0.31 & 0.54 & 0.61 & 0.69 \\
0.05	&0.26	&0.49	&0.55	&0.63	&& 0.25	& 0.49 & 0.55 & 0.64 \\
            \hline\hline
        
        \end{tabular}
    \end{center}

\end{table}

\begin{figure}[!htpb]
    \centering
    \includegraphics{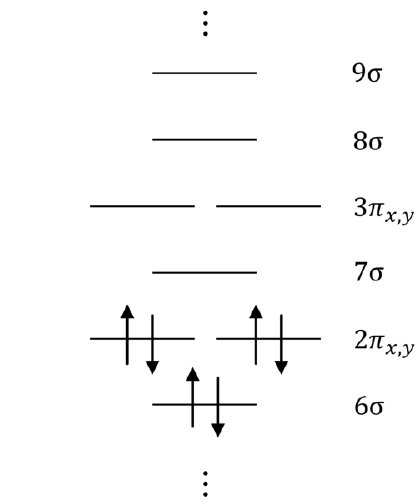}
    \caption{Molecular orbital energy diagram for MgF$^+$.}
    \label{FIG:MGF+_CONFIGURATION}
\end{figure}

Having established that EOM-EA-QED-CCSD-1 can recover electron affinities derived from $\Delta$QED-CC calculations, we now use EOM-EA-QED-CCSD-1 to explore potential energy curves for electron-attached states of a molecule embedded within an optical cavity. Specifically, we consider the electronic spectrum of magnesium fluoride (MgF), as evaluated via EOM-EA-QED-CCSD-1 applied to the cation, MgF$^+$. This system represents an ideal playground for exploring EOM-EA-QED-CCSD-1, as it is known\cite{Bartlett95_3629} that (1) the cation is well-described by a restricted Hartree-Fock reference function at its equilibrium geometry, and (2) EOM-EA-CCSD applied to this state yields excitation energies for {\color{black}neutral} MgF that are in excellent agreement with those obtained from EOM-EE-CCSD and also from experiment. 
Table \ref{TAB:MGF} provides the (cavity-free) electronic spectrum for MgF derived from EOM-EA-CCSD and EOM-EE-CCSD (which were carried out using the aug-cc-pVDZ basis set), as well as from experiment.\cite{HerzbergBook} Here, the Mg--F distance (1.752 \AA) corresponds to the experimental equilibrium bond length of neutral MgF.\cite{HerzbergBook} 
At this geometry, the character of each of the states can be understood in terms of electron attachment to the Hartree-Fock configuration for MgF$^+$  depicted in Fig.~\ref{FIG:MGF+_CONFIGURATION}. The main contributions to the three $\Sigma$-symmetry states ($X~{}^2\Sigma$, $B~{}^2\Sigma$, and $C~{}^2\Sigma$) correspond to electron attachment to the 7$\sigma$, 8$\sigma$, and 9$\sigma$ orbitals in Fig.~\ref{FIG:MGF+_CONFIGURATION}, respectively, while the $A~{}^2\Pi$ state corresponds to the addition of an electron to either the 3$\pi_x$ or 3$\pi_y$ orbital.
\begin{table}[!htpb]
    \caption{The first few energy levels (eV) of MgF determined via EOM-EA-CCSD, EOM-EE-CCSD, and experiment.}
    \label{TAB:MGF}
    \begin{center}
        \begin{tabular}{lccc}

            \hline\hline
        
  	      state  & EOM-EA-CCSD & EOM-EE-CCSD & experiment \\	
      	\hline
$X~{}^2\Sigma$ & IP=7.76 & IP=7.77 & ... \\ 
$A~{}^2\Pi$    & 3.41 & 3.42 & 3.45 \\ 
$B~{}^2\Sigma$ & 4.68 & 4.69 & 4.61 \\ 
$C~{}^2\Sigma$ & 5.25 & 5.27 & 5.27 \\ 

            \hline\hline
        
        \end{tabular}
    \end{center}

\end{table}
In lieu of ground-state energies, we report computed ionization potentials (IPs) for MgF, where the EOM-EE-CCSD value actually corresponds to a $\Delta$CCSD-type IP. We note good agreement between the computed IPs. We also note good agreement between EOM-EA-CCSD and EOM-EE-CCSD derived excitation energies for all states. Despite the use of a modest basis set, both sets of computed excitation energies agree well with experimentally-derived values; the largest deviation from experiment that we observe is 0.08 eV, which corresponds to the overestimation of the $X~{}^2\Sigma$ $\to$ $B~{}^2\Sigma$ transition energy by EOM-EE-CCSD. Hence, we can conclude that EOM-EA-CCSD provides a reasonable description of the cavity-free excitation spectrum for this molecule, at least at the equilibrium geometry.

\begin{figure}[!htpb]
    \centering
    \includegraphics{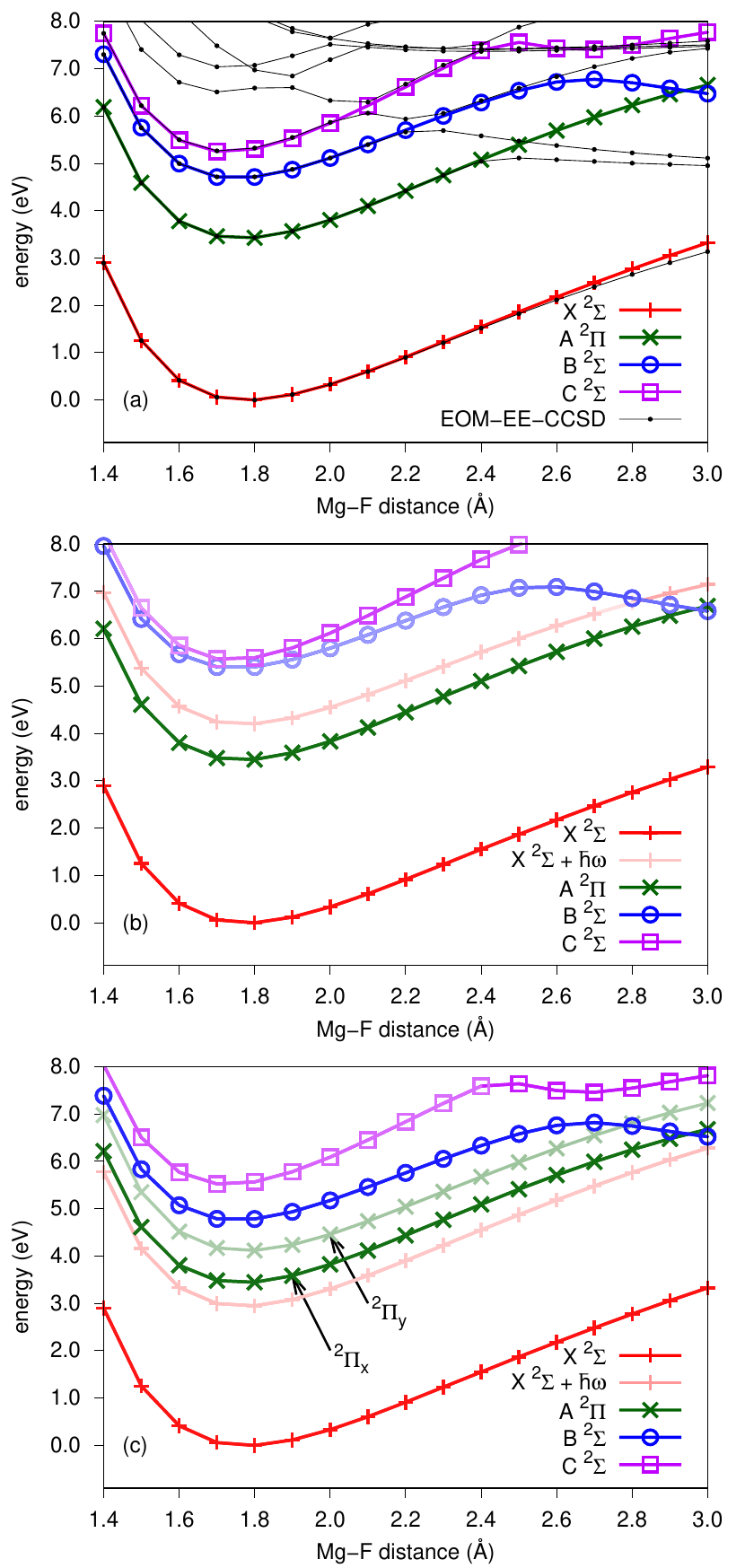}
    \caption{Potential energy curves for MgF (a) in the absence of a cavity, (b) coupled to a cavity mode polarized along the molecular axis with $\lambda$ = 0.05 and $\omega_{\rm cav}$ = 4.7091 eV, and (c) coupled to a cavity mode polarized perpendicular to the molecular axis with $\lambda$ = 0.05 and $\omega_{\rm cav}$ = 3.4262 eV.}
    \label{FIG:MGF}
\end{figure}

Figure \ref{FIG:MGF}(a) depicts potential energy curves for MgF computed at the EOM-EA-CCSD and EOM-EE-CCSD levels of theory, in the absence of a cavity. EOM-EA-CCSD results for the $X~{}^2\Sigma$, $A~{}^2\Pi$, $B~{}^2\Sigma$, and $C~{}^2\Sigma$ curves are provided, as well as multiple EOM-EE-CCSD curves. We find good agreement between EOM-EA-CCSD and EOM-EE-CCSD derived $X~{}^2\Sigma$ curves throughout the entire range of bond lengths considered, while comparable descriptions of the excited-state curves are only obtained up to an Mg--F distance of roughly 2.0--2.4 \AA, depending on the state. There are two primary differences between EOM-EE-CCSD and EOM-EA-CCSD curves at longer bond lengths. First, EOM-EE-CCSD predicts that the energy of the $A~{}^2\Pi$ state decreases beyond 2.4 \AA, while EOM-EA-CCSD predicts that the energy of this state continues to increase for all bond lengths considered. Second, EOM-EE-CCSD predicts a series of avoided crossings in the $\Sigma$-symmetry states at roughly 2.1 and 2.2 \AA~that do not appear in EOM-EA-CCSD derived curves until somewhat larger Mg--F distances (approximately 2.4 and 2.6 \AA). It is difficult to say whether either method is reliable at stretched geometries, as the T1-diagnostic\cite{Schaefer89_81} is quite large for both charge states. At an Mg--F distance of 3.0 \AA, for example, the T1 diagnostics for MgF$^{+}$ and MgF are 0.10 and 0.04, respectively. Nonetheless, EOM-EA-CCSD and EOM-EE-CCSD are in good agreement in the vicinity of the equilibrium geometry, and we believe that EOM-EA-QED-CCSD-1 should be capable of capturing the main qualitative features of the cavity-coupled states in this region.

\begin{figure*}[!htpb]
    \centering
    \includegraphics{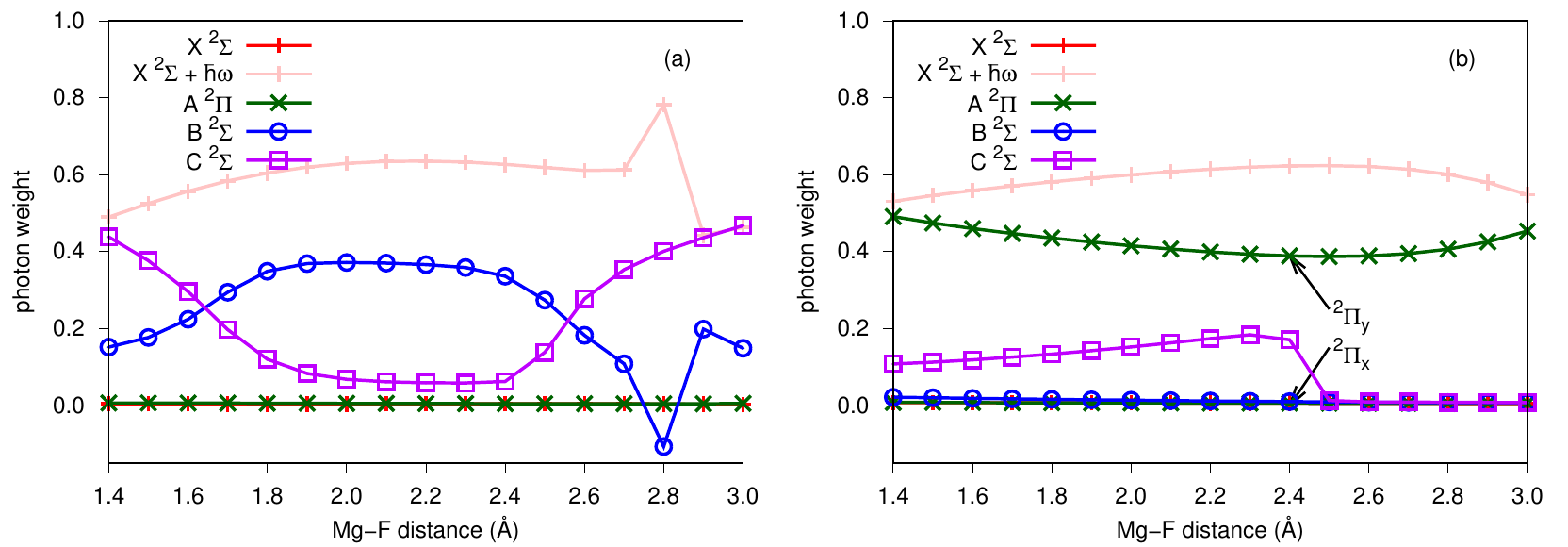}
    \caption{Photon contributions to polaritonic states of MgF coupled to a cavity mode polarized (a) along the molecular axis with $\lambda$ = 0.05 and $\omega_{\rm cav}$ = 4.7091 eV and (b) perpendicular to the molecular axis with $\lambda$ = 0.05 and $\omega_{\rm cav}$ = 3.4262 eV.}
    \label{FIG:MGF_PHOTON_WEIGHT}
\end{figure*}

We now consider MgF coupled to a cavity that supports a single optical mode polarized along the molecular axis [in the $z$-direction, {\em i.e.}, $\bm \lambda$ =  $(0, 0, \lambda)$], with $\lambda$ =  0.05 and $\omega_{\rm cav}$ = 4.7091 eV [Fig.~\ref{FIG:MGF}(b)]. Here, $\omega_{\rm cav}$ has been chosen to match the EOM-EA-CCSD-derived $X~{}^2\Sigma \to$  $B~{}^2\Sigma$ transition energy of isolated MgF at an Mg--F distance of 1.8 \AA. The curves illustrated in Fig.~\ref{FIG:MGF}(b) represent the $X~{}^2\Sigma$, $A~{}^2\Pi$, $B~{}^2\Sigma$, and $C~{}^2\Sigma$ states of cavity-embedded MgF evaluated at the EOM-EA-QED-CCSD-1 level of theory, as well as the state that is dominated by the configuration representing the ground state ($X~{}^2\Sigma$) plus a photon (labeled $X~{}^2\Sigma$ + $\hbar \omega$). These labels have been chosen to reflect the dominant contributions to the wave functions at the equilibrium geometry, and the shading of the curves reflects the photon contributions to each state, with lighter shading indicating greater photon contributions. The principal effect induced by the cavity mode is a pronounced Rabi splitting that reflects strong coupling between the $X~{}^2\Sigma$ + $\hbar \omega$ and $B~{}^2\Sigma$ states; this splitting is 1.20 eV at an Mg--F distance of 1.8 \AA. As a result, the gap between the ground and $B~{}^2\Sigma$ states at this bond length is roughly 0.7 eV larger than that in the case of isolated MgF.

We can quantify the photon character of each state with the relative weights of the photon-containing terms in the wave function:
\begin{equation}
    w_{\rm p} = \frac{\sum_{a} m_a s^a + \frac{1}{2} \sum_{abi} m^{i}_{ab} s_{i}^{ab}}{\sum_{a} (l_a r^a+m_a s^a) + \frac{1}{2} \sum_{abi} (l^{i}_{ab} r_{i}^{ab} + m^{i}_{ab} s_{i}^{ab})}
\end{equation}
These photon weights are presented as a function of Mg--F distance in Fig.~\ref{FIG:MGF_PHOTON_WEIGHT}.
Panel (a) of this figure shows that, near the equilibrium geometry, both the $X~{}^2\Sigma$ + $\hbar \omega$ and $B~{}^2\Sigma$ states have significant photon character, which is consistent with the large Rabi splitting we observe.
At an Mg--F distance of 1.8 \AA, these states have $w_{\rm p}$ = 0.60 and 0.35, respectively, and  the $C~{}^2\Sigma$ state also exhibits non-negligible photon character ($w_{\rm p}$ = 0.12). 
The photon character of the $C~{}^2\Sigma$ state is greater than that of the $B~{}^2\Sigma$ state at short Mg--F distances ($<1.7$ \AA), while the opposite is true over the bond-length range 1.8--2.5 \AA. Beyond 2.5 \AA, the photon weights of these states cross again, before that of the $B~{}^2\Sigma$ state drops precipitously and takes on a negative value at 2.8 \AA.  Meanwhile, the photon character of the $X~{}^2\Sigma$ + $\hbar \omega$ state stays relatively constant with $w_{\rm p} \approx 0.5$--$0.6$, until the spike at 2.8 \AA. These peculiar features in the photon weights of the $X~{}^2\Sigma$ + $\hbar \omega$ and $B~{}^2\Sigma$ states near 2.8 \AA~reflect the proximity to a conical intersection between these states at slightly longer bond lengths. Indeed, we find that the energies of these two states take on complex values, with the real parts being degenerate, for Mg--F distances in the range 2.82 -- 2.84 \AA. This behavior is the hallmark of a defective similarity-transformed Hamiltonian at accidental same-symmetry conical intersections.\cite{Koch17_164105}

Figure \ref{FIG:MGF}(c) illustrates potential energy curves for MgF coupled to an optical mode polarized perpendicular to the molecular axis [$\bm \lambda$ =  $(0, \lambda, 0)$], with $\lambda$ =  0.05 and $\omega_{\rm cav}$ = 3.4262 eV. In this geometry, the $X~{}^2\Sigma$ + $\hbar \omega$ state can potentially couple strongly to one of the $A~{}^2\Pi$ states, so the cavity mode frequency is chosen to match the $X~{}^2\Sigma \to$ $A~{}^2\Pi$ transition energy in isolated MgF (at a bond length of 1.8 \AA). Note that this choice of polarization for the cavity mode breaks the symmetry of the Hamiltonian, but we retain the $\Sigma$ / $\Pi$ labels used in panels (a) and (b) as a matter of convenience. The following cavity-induced changes to the spectrum are observed. First, the cavity mode lifts the degeneracy of the $A~{}^2\Pi$ states, as only one of these states ($A~{}^2\Pi_y$) has the appropriate symmetry to interact with the $X~{}^2\Sigma$ + $\hbar \omega$ state. As a result, two curves labeled $A~{}^2\Pi$ are depicted in Fig.~\ref{FIG:MGF}(c), with the shading indicating the photon character of the wave functions. The $X~{}^2\Sigma \to$ $A~{}^2\Pi_y$ transition energy increases from the isolated-molecule value of 3.4262 eV to 4.1169 eV at an Mg--F distance of 1.8 \AA, whereas the $X~{}^2\Sigma \to$ $A~{}^2\Pi_x$ transition energy is essentially unchanged (3.4478 eV). As in the previous example, a large Rabi splitting between the $X~{}^2\Sigma$ + $\hbar \omega$ and $A~{}^2\Pi_y$ states is observed (1.17 eV), and this pronounced splitting persists over the entire range of Mg--F distances considered. The photon weights provided in Fig.~\ref{FIG:MGF_PHOTON_WEIGHT}(b) are consistent with the large Rabi splittings we observe; both the $X~{}^2\Sigma$ + $\hbar \omega$ and $A~{}^2\Pi_y$ have substantial photon character over all bond lengths. On the other hand, $A~{}^2\Pi_x$, as expected, has essentially zero photon character.

One unexpected feature of the geometry in which the cavity mode is polarized perpendicular to the molecular axis is the apparently non-negligible photon character of the $C~{}^2\Sigma$ state [Fig.~\ref{FIG:MGF_PHOTON_WEIGHT}(b)]. It appears that this character derives from coupling between the original cavity-free $C~{}^2\Sigma$ state and one that is best described as $A~{}^2 \Pi_{y}$ plus a photon.
At an Mg--F distance of 1.8 \AA, a state with $w_{\rm p}=0.45$ lies 1.16 eV higher than $C~{}^2\Sigma$, and the dominant transition amplitude that parametrizes this state confirms that it can be described as $A~{}^2 \Pi_{y}$ plus a photon, ({\em i.e.}, electron attachment to the 3$\pi_y$ orbital of MgF$^+$ plus a photon). Meanwhile, the dominant configuration of the $C~{}^2\Sigma$ represents electron attachment to the 9$\sigma$ orbital, as expected, and the dominant photon-containing contribution to this state looks like electron attachment to 3$\pi_y$ plus a photon. The $C~{}^2\Sigma$ state retains roughly the same photon character until an Mg--F distance of 2.4 \AA, at which point there is an avoided crossing, and the character of the state changes abruptly. This example highlights the usefulness of an {\em ab initio} description of polaritonic structure; the Pauli--Fierz Hamiltonian naturally captures strong coupling scenarios -- and qualitative changes to those couplings -- that could inadvertently be neglected when constructing a model Hamiltonian.

{\color{black}Lastly, we note that the photon character of the cavity-coupled ground state is small at all geometries, for both choices of the cavity mode polarization axis.  As stated above, we observe Rabi splittings as large as 1.2 eV, where the coupling strength can safely be described as ultra-strong. It has been suggested that the ground state can acquire photon character in such cases,\cite{Rubio17_3026} but we find that $w_{\rm p}$ never exceeds 0.008 in any of our calculations. }

\section{Conclusions}

In this work, we have generalized the electron attachment equation-of-motion coupled-cluster approach to the case of strong light-matter coupling via the framework of cavity quantum electrodynamics. Benchmark calculations confirm that the resulting EOM-EA-QED-CCSD-1 formalism yields electron affinities in good agreement with those obtained from $\Delta$QED-CCSD-1-type calculations with an important caveat: the coherent-state basis transformation employed within EOM-EA-QED-CCSD-1 should be defined with respect to the QED-HF dipole moment for the ($N$+1)-electron state, rather than that for the $N$-electron state. EOM-EA-QED-CCSD-1 was also applied to the electronic spectrum of magnesium fluoride, as described via electron attached states starting from the corresponding cation, MgF$^{+}$. We observed strong photon-electron coupling for cavity mode polarizations both parallel and perpendicular to the molecular axis, with the principal effect being a substantial Rabi splitting between a low-lying electronically excited state and a state that resembles the ground electronic state plus a photon. Additional interesting features include  the introduction of an accidental same-symmetry conical intersection and non-negligible coupling between a higher-lying electron/photon states. These observations highlight the value of an {\em ab initio} approach to polaritonic structure.

To conclude, we have established EOM-QED-CC theory as a viable framework for exploring strong electron-photon coupling in particle-non-conserving sectors of Fock space. In addition to the electron attachment approach explored herein, EOM-QED-CC theories could be developed as generalizations of any of the many flavors of standard EOM-CC theory, including the ionization potential,\cite{Snijders92_55,Snijders93_15,Gauss94_8938,Bartlett03_1128,Wloch05_134113,Piecuch06_234107,Krylov11_6028,Krylov12_2726} spin-flip,\cite{Krylov01_375,Krylov04_175} double ionization potential,\cite{Nooijen02_65,Stanton03_42,Bartlett11_114108,Piecuch13_194102,Bartlett17_184101} or double electron attachment\cite{Piecuch13_194102,Kucharski14_114107,Piecuch17_3469,Bartlett17_184101} approaches. Such methods preserve the simplicity of single-reference quantum chemical approaches and yet are often capable of describing the complex electronic structure of highly multireference systems. Hence cavity-QED-based extensions of these methods could enable reliable simultaneous descriptions of both strong electron-electron and strong electron-photon correlation effects. However, it should be stressed that, as we found for the case of electron attachment, accurate results will require that the coherent-state basis be defined according to a QED-HF configuration representative of the target manifold of states, as opposed to that for the reference state.

\vspace{0.5cm}

{\bf Supporting Information} EOM-EE-CCSD, EOM-EA-CCSD, and EOM-EA-QED-CCSD-1 energies for MgF in the aug-cc-pVDZ basis.

\vspace{0.5cm}

\begin{acknowledgments}This material is based upon work supported by the National Science Foundation under Grants No.~CHE-2100984 and CHE-1554354.\\ 
\end{acknowledgments}

\noindent {\bf DATA AVAILABILITY}\\

    The data that support the findings of this study are available from the corresponding author upon reasonable request.

\bibliography{main.bib}

\end{document}